\documentclass[showpacs]{revtex4}
\usepackage{amsmath}
\usepackage{amssymb}
\usepackage{graphicx}

\begin{document}

\title{Viscous dark fluid}

\author{Vladimir Folomeev}
\email{vfolomeev@mail.ru} \affiliation{Institute of Physics of NAS
KR, 265 a, Chui str., Bishkek, 720071,  Kyrgyz Republic}

\author{Victor Gurovich}
\email{tsalevich@yahoo.com} \affiliation{Physics Department,
Technion, Technion-city, Haifa 32000, Israel}

\begin{abstract}
The unified dark energy and dark matter model within the framework of a model
of a continuous medium with bulk viscosity (dark fluid) is considered.
It is supposed that the bulk viscosity coefficient is an arbitrary function of the Hubble
parameter. The choice of this function is carried out under the
requirement to satisfy the observational data from recombination ($z\approx 1000$) till present time.
\end{abstract}

\pacs{95.36.+x, 98.80.Es}
\maketitle
Keywords: Dark energy, dark matter, viscous fluid.

\section{Introduction}
Modelling of an accelerated expansion of the present Universe lies on the way
of creation of phenomenological models which may explain the
observational data on one set of parameters and compare them with predictions
of the models on other set. For example, a theoretical model adapts for the correct
description of the acceleration of the Universe in an accessible interval $z$. Further,
the results of modelling are extrapolated for large $z$ which are not accessible for observations yet.
The corresponding cosmological scenario defines growth of the large-scale structure which determines the present-day
fluctuations of the microwave background radiation.  Certainly, the models under consideration should
not contradict the available observational data within the framework of general relativity in a field of its applicability. For the
specified purposes a number of cosmological models successfully applied in the past in
the theory of the early Universe is used. These are cosmological models with various
scalar and non-scalar fields filling the space together with cold dark matter (see, e.g., the reviews \cite{Sahni}).
Cosmological models of the present accelerated Universe within the framework of high-order theories of gravity (HOTG)
are also quite popular \cite{Carrol}.

A number of models have recently been suggested \cite{Ren, Brevik,Odin} which describe the present
Universe with use of models of a continuous medium in the presence of bulk viscosity.
Consideration of effects of viscosity within the framework of HOTG was also carried out
\cite{Brevik1}. Note that such models were well-known in the theory
of the early Universe (see, for example, \cite{Murphy,Barrow}).
In particular, in Ref. \cite{Barrow} a few exact solutions with the constant bulk viscosity
coefficient and with the bulk viscosity being an arbitrary power function of energy density were obtained.
In Ref. \cite{Ren} the model of viscous
dark fluid is considered. The main result of this paper is the model with the constant bulk viscosity coefficient.
The model fits the observational data on luminosity at an acceptable level. In Ref. \cite{Brevik} the models both
with the constant bulk viscosity coefficient and the bulk viscosity linearly proportional to the
Hubble parameter are examined. The question about influence of viscosity on presence of a singularity
in the model in the future (the so-called Big Rip) is investigated.

In this paper we consider a model of "viscous dark fluid" with the bulk viscosity coefficient $\mu(H)$
which  depends on the Hubble parameter arbitrarily.  Unlike Ref. \cite{Ren}, comparison of the model with
the observational data is not restricted to the observational data on luminosity.
The model is being compared with results of observations on change of the deceleration parameter
 $q$ and values of the Hubble parameter in the range $2>z>0$. It will be shown below that
the model with the constant bulk viscosity coefficient does not provide a good description
for $q(z)$ and $H(z)$ which follow from the observations.
We propose such a dependence $\mu(H)$ which is adequate to the mentioned observations.
The proposed model is extrapolated for $z$ beyond the specified range $2>z>0$.

\section{Equations and solutions}
The metric of the flat Universe is taken as:
\begin{equation}
\label{metr}
ds^2=c^2 d t^2-a(t)^2(dx^2+dy^2+dz^2).
\end{equation}
The corresponding 0-0 component of the Einstein equations is:
\begin{equation}
\label{00}
H^2=\frac{1}{a^2}\left(\frac{d a}{d t}\right)^2=\frac{8\pi G}{3 c^2}\varepsilon,
\end{equation}
where $H$ is the Hubble parameter, $\varepsilon$ is the energy density of matter. Introducing the
dimensionless energy density
$$
\delta=\frac{\varepsilon}{\varepsilon_*},
$$
where the critical density $\varepsilon_*=3 c^2 H_0^2/8\pi G$ (the subscript 0 indicates the value
of the parameter in the present time), one has from \eqref{00}:
\begin{equation}
\label{00_n}
\delta=\frac{H^2}{H_0^2}=h^2.
\end{equation}
Here the dimensionless Hubble parameter $h$ is expressed in units of its present value $H_0$.

The corresponding energy conservation law  for the viscous dark fluid can be obtained from the
equation:
\begin{equation}
\label{cons}
\left[T^k_i+\tau^k_i\right]_{;k}=0,
\end{equation}
where the energy-momentum tensor of matter is
$$T^k_i=(\varepsilon+p)u_i u^k-\delta_i^k p,$$
and the tensor of viscosity is
$$\tau^k_i=\mu u^l_{;l} (\delta^k_i-u_i u^k)$$
with the bulk viscosity coefficient $\mu$.
By carrying out covariant differentiation in
\eqref{cons} with taking into account the metric
\eqref{metr}, one can obtain the following equation:
\begin{equation}
\label{energ}
\frac{d\delta}{d\theta}+3h\delta=9\lambda h^2.
\end{equation}
Here the dimensionless time
$\theta=H_0 t$ is introduced, and the bulk viscosity coefficient redefined with help of the
dimensionless parameter $\lambda$ as
$$\mu=\varepsilon_* \lambda/H_0.$$

Similarly to \cite{Ren}, let us suppose that the viscous medium has the pressure $p=0$.
Then equations \eqref{00_n} and \eqref{energ} imply one equation for
the dimensionless Hubble parameter $h$.
The case $\lambda=const$ is equivalent to the Murphy's model
\cite{Murphy}. As a matter of fact this model was used in \cite{Ren} for comparison with the
observations. The models of the present Universe mentioned in Introduction use the fact that
the deceleration parameter $q$ in the past at $z\gg 1$ was close to $q\approx 0.5$ (cold dark matter), and
in the present time the Universe expands with acceleration.  For this reason all the models (both
with scalar fields and in HOTG) are created in such a way that the present inflation appears for rather
small values of the Hubble parameter, or what is the same thing, for small average density of matter.
In this connection we
will consider the model \eqref{energ} in which the dimensionless bulk viscosity of the dark fluid $\lambda$ is
not a constant but an arbitrary function of the parameter $h$. In this paper we chose this function as:
\begin{equation}
\label{visc}
9\lambda=3\tanh{\left(\frac{b}{h^n}\right)},
\end{equation}
where $b,n$ are arbitrary constants which will be defined from the observational data later.

For comparison with the observational data, it is convenient to rewrite equations \eqref{00_n} and \eqref{energ}
through the redshift $z=1/a-1$. Then one has the equation for $h$ in the form:
\begin{equation}
\label{eq_h}
-2(z+1)\frac{dh}{dz}+3h=3\tanh{\left(\frac{b}{h^n}\right)},
\end{equation}
and the deceleration parameter $q$ will be:
$$
q=-\left(1-\frac{(z+1)}{h}\frac{dh}{dz}\right).
$$

\begin{figure}[ht]
\begin{center}

  \includegraphics[width=12cm]{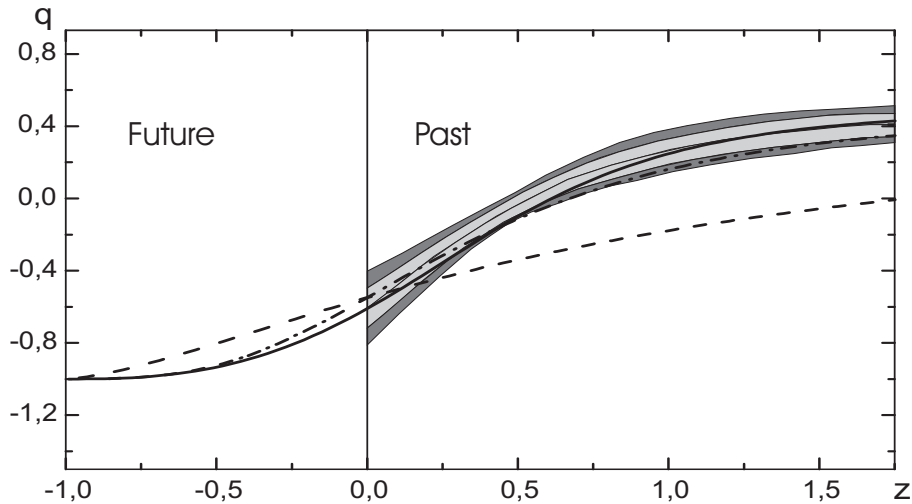}
 \caption{ The deceleration parameter for our model (thick solid line), the model with
the constant bulk viscosity from \cite{Ren} (dashed line), the $\Lambda$CDM model (dashed-dot line).
The central solid thin line represents the best-fit, the light grey contours
represent the $1\sigma$ confidence level, and the dark grey contours represent the $2\sigma$
confidence levels (the data taken from \cite{sahni1}).
}
\label{decel}
\end{center}
\end{figure}

The parameters $b,n$ in the model are being chosen from the requirement  that in the present time
(at $h=1$ with account of selected normalization) the deceleration parameter $q$ should be close to the
observable value $q\approx -0.6$. This value of $q$ can be obtained at $b=0.95$  and $n=2$. Using these
parameters, we have compared the model under consideration with the observational data in the range $2>z>0$.
The results are presented in Fig.~\eqref{decel}. For comparison with our model, the $\Lambda$CDM model with
the same ''initial conditions`` is also shown in Fig.~\eqref{decel}. The model with
the constant bulk viscosity from \cite{Ren} is presented as well.

From Fig.~\eqref{decel}, it follows that the model with the constant bulk viscosity deviates from the
observations appreciably. The model under consideration with $\lambda(h)$ lies within the confidence
levels. Note that this model close to the $\Lambda$CDM model (with $\Omega_m=0.3$ and $\Omega_{\Lambda}=0.7$),
although there is no any special $\Lambda$-term in the model. Outside the interval of observations (in practice at
$z \gtrsim 2$) our model, the $\Lambda$CDM model and extrapolation of the observable data from \cite{sahni1}
are close to each other and give  $q\approx 0.5$. The model with constant $\lambda$ has much less value of $q$ and
reaches $q\approx 0.5$ at $z\geq 40$.

An extrapolation in the future for all three models is also shown in Fig.~\eqref{decel}. Asymptotically
($z\rightarrow -1$) all three models tend to the de Sitter model.

\section{Conclusion}

The bulk viscosity in our model is an example of a dynamic $\Lambda$-term. However, our
model is close to the $\Lambda$CDM model.  As it was rightly noted in \cite{Ren}, the
model with the viscosity does not give possibility to divide the true dust filling the Universe,
and dark matter generated by the bulk viscosity. That is why it is difficult to introduce a phenomenological equation
of state $p=w \varepsilon$ which is often used for interpretation of the observational data.
Influence of the bulk
viscosity on formation of the large-scale structure of the Universe demands special examination. But taking
into account that the viscosity is being ''involved`` at rather small $z\approx 2$, it is possible to expect its influence on
the dynamics of galactic clusters only at later non-linear stage.

\end{document}